\documentclass[%
 preprint,
nofootinbib,
 amsmath,amssymb,superscriptaddress
]{revtex4-1}

\usepackage{graphicx}% Include figure files
\usepackage{dcolumn}% Align table columns on decimal point
\usepackage{bm}% bold math
\usepackage{hyperref}% add hypertext capabilities
\usepackage{amsmath} %for subequations
\usepackage{color}
\usepackage[dvipsnames]{xcolor}
\hypersetup{
	colorlinks,
	linkcolor={red!95!blue},
	citecolor={Cerulean!100},
	urlcolor={blue!80!black}
} %hyperlink colors

\begin{document}

\preprint{\href{http://www.nature.com/articles/srep24283}{Submitted to Scientific Reports}}

\title{Theory of Thomson scattering in inhomogeneous media}% Force line breaks with \\

\author{P. M. Kozlowski}
\email{Pawel.Kozlowski@physics.ox.ac.uk}
\affiliation{%
 Department of Physics, University of Oxford, Parks Road, Oxford, OX1 3PU, United Kingdom
 %Clarendon Laboratory,University of Oxford, South Parks Rd., Oxford, OX1 3PU, United Kingdom
}

\author{B. J. B. Crowley}
\affiliation{%
 Department of Physics, University of Oxford, Parks Road, Oxford, OX1 3PU, United Kingdom
 %Clarendon Laboratory,University of Oxford, South Parks Rd., Oxford, OX1 3PU, United Kingdom
}
\affiliation{AWE plc, Reading RG7 4PR, United Kingdom}

\author{D. O. Gericke}
\affiliation{Centre for Fusion, Space and Astrophysics, Department of Physics, University of Warwick, Coventry CV4 7AL, United Kingdom \\}

\author{S. P. Regan}
\affiliation{Laboratory for Laser Energetics, University of Rochester, 250 East River Road,
Rochester, NY 14623, United States \\}

\author{G. Gregori}%
\affiliation{%
 Department of Physics, University of Oxford, Parks Road, Oxford, OX1 3PU, United Kingdom
 %Clarendon Laboratory,University of Oxford, South Parks Rd., Oxford, OX1 3PU, United Kingdom
}

\date{\today}% It is always \today, today,

\begin{abstract}
\bf Thomson scattering of laser light is one of the most fundamental diagnostics of plasma density, temperature and magnetic fields. It relies on the assumption that the properties in the probed volume are homogeneous and constant during the probing time. On the other hand, laboratory plasmas are seldom uniform and homogeneous on the temporal and spatial dimensions over which data is collected. This is particularly true for laser-produced high-energy-density matter, which often exhibits steep gradients in temperature, density and pressure, on a scale determined by the laser focus. Here, we discuss the modification of the cross section for Thomson scattering in fully-ionized media exhibiting steep spatial inhomogeneities and/or fast temporal fluctuations. We show that the predicted Thomson scattering spectra are greatly altered compared to the uniform case, and may even lead to violations of detailed balance. Therefore, careful interpretation of the spectra is necessary for spatially or temporally inhomogeneous systems. 
%\begin{description}
%\end{description}
\end{abstract}

\maketitle

%\tableofcontents

%\section{\label{sec:level1} Introduction\protect }

%This, when applied to inertial fusion capsules \cite{glenzer1,hurricane}, significantly complicates the determination of the properties of the dense, compressed core. 

%X-ray scattering experiments using $4^{th}$ generation light sources are especially vulnerable due to very small samples and probing on the femto-second time scales.

Fourth generation light sources hold the promise of improving our understanding of extreme states of matter by providing a probe which can penetrate through the enormous densities produced in inertial confinement fusion or laboratory astrophysics experiments. Thomson scattering by free electrons has emerged as a powerful diagnostic for such systems, through its extension from the optical through the x-ray regime, allowing it to take full advantage of the new capabilities provided by fourth generation light sources (e.g. LCLS, SACLA, European XFEL, SwissFEL) \cite{sheffield, pile, emma1}. Thomson scattering allows for the measurement of density, temperature, and ionization state in plasmas, leading to an effective characterization of the plasma equilibrium state \cite{evans, glenzer1, sheffield}, and progress in the understanding of the properties of high-energy-density matter has significantly relied upon this technique \cite{gregori1,glenzer2,glenzer3,saiz,kritcher,brown,chapman15}.
Thomson scattering has also been utilized for probing temperatures and magnetic fields in tokamaks \cite{peacock1969,carolan1978}.

Investigations using Thomson scattering to date have been based on the assumption of a homogenous or weakly inhomogeneous plasma. This limitation becomes particularly restrictive when considering the ultra-short x-ray pulses and near diffraction limited laser spot sizes of fourth generation light sources \cite{nagler, emma1}. Under such conditions, large spatial and temporal gradients in the properties of matter are not negligible and are mainly determined by the extent of laser focus \cite{drake}. This situation is exemplified by inertial fusion experiments, where
the capsule is compressed by a series of shocks \cite{glenzer1,hurricane}, which introduce inhomogeneities on the scale of the particle mean free path ({\it i.e.}, the shock width) and this significantly complicates the determination of the properties of the dense, compressed core
in a scattering experiment \cite{regan}. Interpreting the scattering signals with models developed for homogenous equilibrium systems may thus lead to significant errors in the inferred properties of the matter. 

Here we develop a generalization of the theory for Thomson scattering to spatially inhomogeneous systems, with a scale-length of the gradient comparable to the scattering wavelength, as well as to rapidly varying plasma conditions. In order to capture the fast externally driven relaxation processes in a sample, ultra-short pulses with durations on the order of the inverse of the plasma frequency are needed. We demonstrate that, under these conditions, both strong density and temporal gradients result in significant changes in the predicted spectra, modifying the intensity and width of the inelastic scattering peaks due to collective electron excitations (plasmons). As Thomson scattering diagnostics yield results by matching theoretical predictions with experimental data, these modifications have important consequences when interpreting experimental data.

%\section{\label{sec:level1} Theory \protect }
\subsection*{\label{sec:level2} Limitations of theories for homogeneous systems}

We consider a scattering probe of wavelength $\lambda_0$. The wavenumber of the transferred momentum is determined by the scattering geometry as $k \equiv |{\bf k}| \approx (4\pi/\lambda_0) \sin(\theta/2)$, where $\theta$ is the scattering angle. The spatial scale-length sampled by the scattering probe is thus $\lambda_p \sim 1/k$. The scale length of microscopic density fluctuations in equilibrium plasmas is determined by the screening length $\lambda_{scr}$ of electric fields \cite{evans,sheffield}. This length is usually taken to be the Debye length in a classical plasma or the Thomas-Fermi screening length in a degenerate electron gas. Whether the single particle behaviour or collective excitations are probed (noncollective versus collective scattering) depends on the ratio of these scale length. Thus, the scattering condition is usually cast by the parameter $\alpha = \lambda_p/\lambda_{scr}$.

Noncollective scattering corresponds to $\alpha \ll 1$. In this case, the total scattering signal is the incoherent sum of individual scatterers. Even in an inhomogeneous system, different regions in the plasma will scatter an amount of photons proportional to the local conditions, and the resultant spectrum is expected to be the sum of these local contributions. In the opposite case of collective scattering with $\alpha>1$, the picture is qualitatively different: now the spatial arrangements of free charges within a distance $\lambda_p$ give rise to a coherent superposition of scattering waves. Since plasmas support electrostatic waves (plasmons), the scattered photons gain or lose an amount of energy close to $\hbar \omega_{pe}$, where $\omega_{pe}$ is the plasma frequency \cite{glenzer1}. Plasmon resonances appear in the scattering spectrum as a result of these wave-particle interactions.

Thomson scattering probes the fluctuations of a system. In equilibrium systems, these can be either microscopic density perturbations due to bound electrons or screening clouds, or collective excitations such as plasmons. Likewise, hydrodynamic gradients in the plasma conditions may also result in light scattering if they occur on the scale of the probe volume. Moreover, the dispersion of the plasma waves may be affected by gradients. If the scale length of macroscopic spatial inhomogeneities, $\Lambda$, is much larger than the length scale probed, that is $\lambda_p$, we can still apply the equilibrium theory locally, {\it i.e.}, on scales $\sim$$\lambda_p$. Hence, the resultant spectrum is simply the sum of the weighted contributions of different regions in the plasma. For strong gradients with $\Lambda \lesssim \lambda_p$, the plasmon dispersion is changed and, thus, the coherence relation between scatterers is altered. Consequently, macroscopic spatial gradients must be considered in the theoretical description explicitly (see also Ref.~\cite{gregori_pre1999} for further considerations). Similar difficulties arise if the duration of the scattering probe, $\tau$, is shorter than the time required for screening to be established and the system is dynamically evolving in states far from equilibrium. The typical time scale for such processes is on the order of 
 $2\pi/\omega_{pe}$.

\section*{Results} 
\subsection*{\label{sec:level2}  Theory for inhomogeneous systems}

The differential cross section of Thomson scattering is determined by the dynamic structure factor $S({\bf k},\omega)$, where $\hbar {\bf k}$ is the change in photon momentum and $\hbar \omega$ is the energy gain (or loss) of the photon during scattering \cite{glenzer1}. In an isotropic medium all directions are equivalent and the dependence on the modulus of the wavenumber is sufficient, that is $S(k,\omega)$. The structure factor is the Fourier transform of the electron density-density correlation function \cite{ichimaru,crowley}, and essentially an extension of the well-known form factor used in x-ray crystallography \cite{kittel}. Thus, it contains all electron correlations and their dynamics including collective excitations in the system. A number of different sources may contribute to the total dynamic structure factor and thus to the Thomson scattering signal. The total dynamic structure factor is usually written as \cite{chihara}:

\begin{equation}\label{eq:ChiharaEq}
S\left(k, \omega \right) = \left| f_{I}(k) + q(k) \right| ^{2} S_{ii} \left(k,\omega \right) + Z_f S_{ee}^{0}(k, \omega) + Z_c \int \tilde{S}_{ce} \left(k,\omega - \omega ' \right) S_s(k, \omega' ) d\omega ' \,.
\end{equation}
The first term deals with electrons which follow the motion of the ions. These motions are described by the ion-ion density correlation function $S_{ii}(k,\omega)$, with
$f_{I}(k)$ the scattering form factor of the bound electrons in the ion and $q(k)$ the contribution arising from the screening cloud around an ion.
The second term in Eq. (\ref{eq:ChiharaEq}) represents free electrons, those which do not follow the ion motions. The last term consists of inelastic scattering by core electrons, where $\tilde{S}_{ce} \left(k,\omega \right)$ is the structure factor of the core electrons inside an ion and $S_s(k, \omega)$ the self structure of the ions, which describes their thermal motion. Here, $Z_f$ and
$Z_c$ are the number of free electrons and core electrons per ion, respectively.

In principle, gradients correction will affect all three terms in Eq.~(\ref{eq:ChiharaEq}). However, we expect the largest modifications to occur for the free electron feature, i.e., $S_{ee}^{0}(k, \omega)$. The other terms are strongly related to the microscopic bound states that will be unchanged by the gradients in the plasma environment. The screening clouds will also be modified only slightly due to hydrodynamic gradients. Thus, we focus here on the effect of collective electron excitations and their dispersion due to gradients in the plasma.

Spatial gradients and fast relaxation processes, including the build-up of correlations and screening, can be described on the basis of the Kadanoff-Baym equations \cite{KB_book}. Although numerical solutions are feasible \cite{KB_00}, they are restricted for easy situations and simple approximations for the interactions. However, a more practical way is possible for weakly coupled plasmas of interest. Taking $S(k,\omega) \equiv S_{ee}^{0}(k, \omega)$, the approximate dynamic structure factor of free electrons in a weakly coupled plasmas is given by the dielectric superposition principle \cite{ichimaru}
\begin{equation} \label{eq:DSFee}
S({\bf k},\omega) = \frac{S^{\rm id}{({\bf k},\omega)}}{|\epsilon \left({\bf k},\omega\right)|^2} \,,
\end{equation}
where $S^{\rm id}{({\bf k},\omega)}$ is the dynamic structure factor for an ideal (noninteracting) gas, and $\epsilon ({\bf k},\omega)$ is the dielectric (screening) function in the random phase approximation (RPA). The latter is given by $\epsilon \left({\bf k},\omega\right) = 1 + \chi^0 \left({\bf k},\omega\right)$, where $\chi^0 \left({\bf k},\omega\right)$ is the susceptibility of the ideal Coulomb plasma.
Eq.~\eqref{eq:DSFee} is an approximation in first order with respect to the correlation strength in the plasma. Although strong coupling effects may need to be taken into account under certain conditions, in most situations concerning laboratory experiments the electrons are nearly or fully degenerate and therefore weakly coupled. The applicability of RPA to first order is often justified for relatively uniform systems, and higher order corrections to the RPA are typically small. Static and dynamic local field corrections act to slightly modify the dispersion relation (i.e., the position of the plasma wave resonance peaks of the structure factor) and can in principle be accounted for within the same theoretical scheme discussed in this paper. Eq.~\eqref{eq:DSFee} also returns to the familiar fluctuation-dissipation theorem (FDT) for systems in local thermodynamic equilibrium (LTE) \cite{ichimaru}. For systems which are not in LTE different approaches may be considered for the structure factors \cite{gregori_pre2006, gregori_hedp2007}; however, in the context of random phase approximation, the dielectric superposition principle is exact and we may proceed with a description of the dielectric function for systems considerably departing from homogeneity and equilibrium. 

The direct application of Eq.~\eqref{eq:DSFee} is crucial for systems with spatial and temporal gradients as the usual application of the FDT limits the theory to the LTE regime. To extend the applicability of the theory to inhomogeneous plasmas, we follow the approach given by Belyi \cite{belyi} and Bornatici and Kravtsov \cite{bornatici}. Assuming the susceptibility to be a smooth function of macroscopic space ($\bf r$)  and time (t)
coordinates, one may apply a first order gradient expansion in the microscopic variables
\begin{equation}
\chi \left({\bf k},\omega\right) \approx \chi^{\rm eq}\left({\bf k},\omega\right) - i \dfrac{\partial}{\partial \bf{k}} \cdot \dfrac{\partial}{\partial \bf{r}} \chi^{\rm eq}\left({\bf k},\omega\right) + i \dfrac{\partial}{\partial \omega} \dfrac{\partial}{\partial t} \chi^{\rm eq}\left({\bf k},\omega\right) \,.
\end{equation}
Here, the index `eq' labels the susceptibility for homogeneous systems in thermodynamic equilibrium.

Introducing a scale for the gradients in space and time, the above form can further be simplified: $\partial/\partial {\bf r} \approx 1/{\Lambda}$ and $\partial/\partial t \approx 1/{\tau}$. Here, $\Lambda$ represents the spatial gradient along the direction of the scattering wavenumber $\bf k$ where  the sign indicates increasing or decreasing plasma parameters. Unless the spatial gradients are distributed uniformly along all directions, as for example in isotropic turbulence, the scattering spectrum will depend on the specific geometry. Similarly, the time constant $\tau$ gives either the strength of externally driven changes ({\it e.g.}, heating) or the relaxation time within the electron system. 
 
The dielectric response of the system can then be constructed in the usual way
\begin{equation}
\epsilon({\bf k},\omega) \approx 1+ \chi^{\rm eq}\left({\bf k},\omega\right) - i \dfrac{1}{\Lambda} \dfrac{\partial}{\partial k} \chi^{\rm eq}\left({\bf k},\omega\right) + i \dfrac{1}{\tau} \dfrac{\partial}{\partial \omega} \chi^{\rm eq}\left({\bf k},\omega\right) \,.
\label{eq:dk_exp}
\end{equation}
For noncollective scattering, $\chi\left({\bf k},\omega\right) \ll 1$,  gradients have no significant effect on the dielectric response. On the other hand, in collective scattering the susceptibility cannot be neglected, and $\epsilon({\bf k},\omega) \sim \epsilon^{\rm eq}({\bf k},\omega)$ follows only if 
$k \Lambda \sim \Lambda/\lambda_p \gg 1$ and $\omega \tau \sim \tau /\tau_p \gg 1$, as discussed before.

The above expansion can be understood from the constitutive relation between the displacement, $\bf D$, and the electric $\bf E$, fields in a macroscopic medium \cite{bornatici}: 
\begin{equation}
D_{i}({\bf r},t) = \int d^{3} r' \int\limits_{-\infty}^{t} dt' \epsilon_{ij}({\bf r},t;{\bf r}',t')E_{j}({\bf r}',t'),
\label{eq:Di}
\end{equation}
where $\bf r$ and $t$ are the space and time coordinates, respectively.
This implies that the dielectric tensor is not a function only of ${\bf r}-{\bf r}'$ and $t-t'$, as we would have in a uniform and stationary medium, but instead it is a function of the space and time variables $({\bf r},t)$ and $({\bf r}',t')$ independently. Through a change of variables the dielectric function may be expressed as

\begin{equation}
\epsilon_{ij}({\bf r},t;{\bf r}',t') \to \epsilon_{ij}\left({\bf r}-{\bf r}',t-t'; \mu {\bf r}', \mu t'\right),
\end{equation}
where the "fast" variables $\Delta \bf{r} = {\bf r}-{\bf r}'$ and $\Delta t = t-t'$ are associated with the microscopic spatial ($\bf k$) and temporal ($\omega$) dispersion in Fourier space. The remaining "slow" variables, $\mu {\bf r}$ and
$\mu t$, account for the macroscopic inhomogeneities of the medium.
The factor $\mu \approx \Lambda/\lambda_p\, ;\, \tau/\tau_p$ compares the scales of the fast and slow variables.
When $\mu \gtrsim 1$, the dielectric function is expanded in Taylor series with respect to $\mu$:
\begin{equation}
	\epsilon_{ij}({\bf r},t;{\bf r}',t') \approx \epsilon_{ij} + \mu\Delta {\bf r} \cdot \frac{\partial \epsilon_{ij}}{\partial {\mu \bf r}}  + \mu \Delta t \frac{\partial \epsilon_{ij}}{\partial \mu t},
\end{equation}
and assuming that the electric field is locally of an eikonal form, $\exp[ i \left( {\bf k} \cdot \Delta {\bf r} - \omega \Delta t \right)]$,
Eq.~(\ref{eq:dk_exp}) is then retrieved \cite{belyi,bornatici}. From the given dielectric function, the free electron dynamic structure factor of inhomogeneous plasmas can be obtained from Eq.~\eqref{eq:DSFee}. As the ideal part is, of course, unchanged, the contribution of gradients to the Thomson scattering signal are all contained in the expression for the dielectric constant Eq.~\eqref{eq:dk_exp}. Higher-order terms in the Taylor expansion are neglected. Accordingly, our theory presents the first-order correction to the equilibrium treatment and is thus limited to moderate gradients. Of course, in real experiments this condition may not apply and higher-order terms should be considered as they may yield details of the complex relation between the the electric and density fluctuations implicit in the non-local Poisson equation. In the latter case, our theory will give at least a clear signal that an equilibrium treatment is not applicable. It should be noted here that the above result has been derived upon the assumption of geometric optics, thus this method of addressing inhomogeneities is applicable not just to the free electron dynamic structure factor, but any calculation of the dynamic structure factor which utilizes the dielectric susceptibility. This fact is indeed important if we intend to apply the same analysis to the other terms in Eq.~(\ref{eq:ChiharaEq}). The main difference in implementing this formula for terms other than the free electron dynamic structure factor, will arise from whether or not Eq.~(\ref{eq:DSFee}) is applicable.

\subsection*{\label{sec:level2}  Application of the theory}

In order to evaluate the effects of spatial and temporal density variations in the scattering spectrum, we consider a practical example of a fully ionized dense deuterium plasma with an average electron density
$n_{e}= 2.2 \times 10^{23} \: \rm cm^{-3}$ and an electron temperature $T_{e}= 8  \: \rm eV$ (see, {\it e.g.}, Regan {\it et al.} \cite{regan} for the experimental setup). Under such conditions the Fermi energy is comparable to the thermal energy and thus we need to take into account quantum effects by choosing a suitable form for $\chi\left({\bf k},\omega\right)$. For the present example we use the random phase approximation \cite{gregori1,Landau}:
\begin{equation}
\epsilon\left({\bf k},\omega\right) = 1 - \frac{e^2}{\hbar \epsilon_{0} k^2} \int \frac{f(\boldsymbol{p} + \hbar \boldsymbol{k}/2) - f(\boldsymbol{p} - \hbar \boldsymbol{k}/2)}{\boldsymbol{k} \cdot \boldsymbol{p}/m_{e} - \omega - iv} \frac{2 d^3 p}{(2\pi\hbar)^3}
\label{eq:chiLandau}
\end{equation}
where $f(\boldsymbol{p})$ is the electron distribution function.

This plasma is probed with x-rays with an energy of 2,960 eV at a scattering angle of $\theta = 40^{\rm o}$, giving $\alpha= 2.17$, i.e., collective scattering conditions. The characteristic probe scale length is $\lambda_p \approx 0.1$ nm, and the relaxation time is $\tau \approx 0.2$ fs. This suggests that strong spatial gradients may be observable whereas it will be improbable to find suitable ultra-fast probes.

\begin{figure}[p]
\centering
\includegraphics[width=0.9\linewidth]{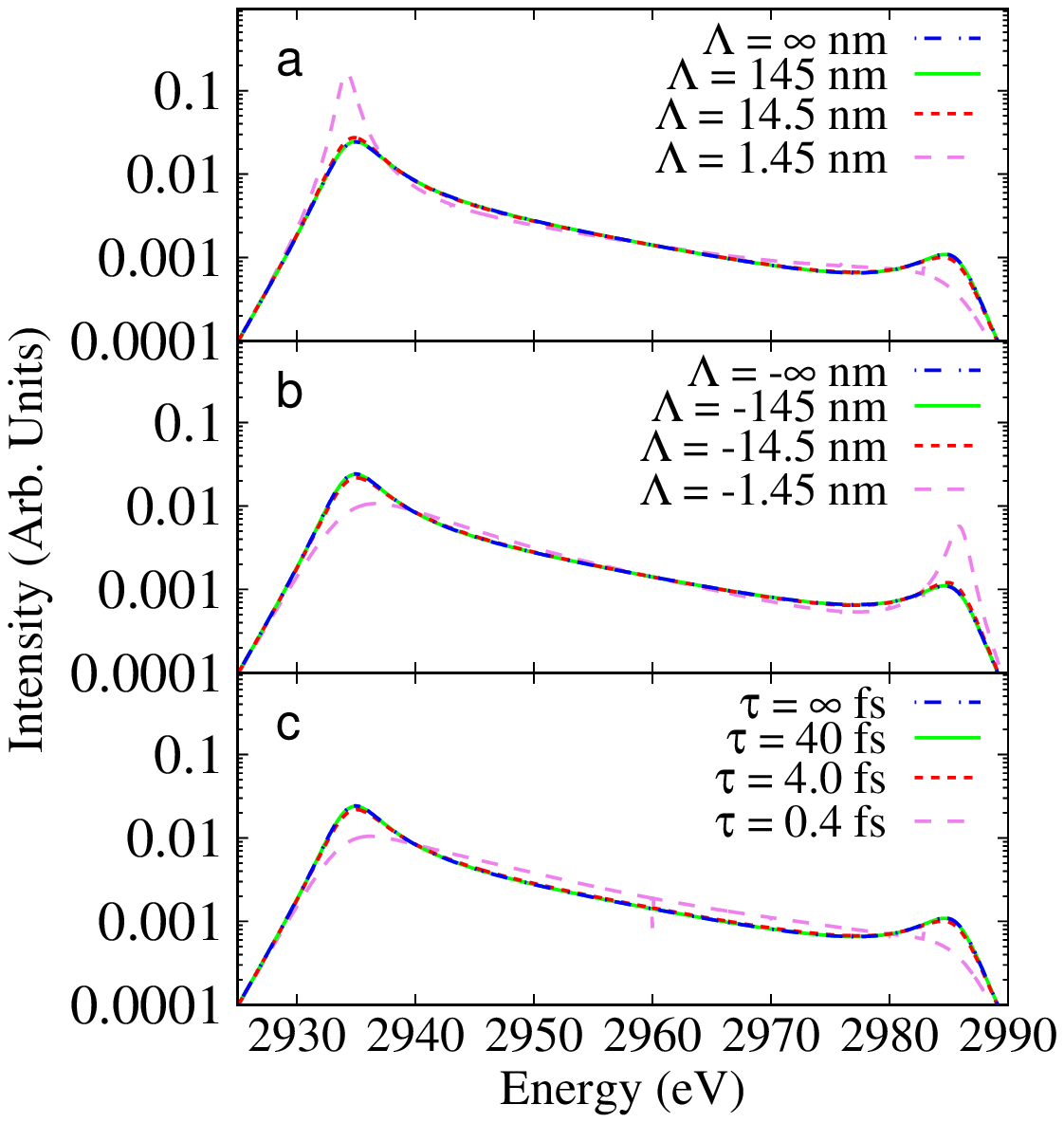}
\caption{{\bf Effects of gradients in the intensity of the plasmon resonances.}
a) Calculation of Thomson scattering intensity, which is proportional to  $S(k,\omega)$, with different values of $\Lambda$ and $\tau=\infty$. The spatial gradients are all assumed to have a component parallel to $\bf k$ where $k = 1.03 \times 10 \: \rm m^{-1}$. b) Same as a) but with the direction of the gradients reversed. c) Calculation of $S(k,\omega)$ with different values of $\tau$ and $\Lambda=\infty$.}
\label{fig:Spectrafordifferentspatialgradients4}
\end{figure}

In Figure~\ref{fig:Spectrafordifferentspatialgradients4}a, we have plotted the dynamic structure factor $S({\bf k},\omega)$ giving the form of the expected Thomson scattering signal for different values for the spatial  scale $\Lambda$ (and taking $\tau \to \infty$). We observe large changes in both the width and relative heights of the blue and red shifted plasmon resonances when $\Lambda/\lambda_p \lesssim 10$. Moreover, the relative intensity of the two resonances changes depending on the direction of the gradients with respect to the probe (parallel or anti-parallel). In Figure~\ref{fig:Spectrafordifferentspatialgradients4}c, we have considered the case of temporal fluctuations comparable to the probe duration (and with no spatial gradients). Again, the predicted dynamic structure factor is significantly broadened for $\tau/\tau_p \lesssim 2$. This effect is to be expected, since a shorter pulse duration would have the same effect as a higher collision frequency between scatterers, which results in a broadening of the plasmon lines.

As expected, the detailed balance relation is not recovered for nonhomogeneous and nonequilibrium systems. 
In a homogeneous plasma, the intensity of the plasmon resonances is determined by detailed balance \cite{ichimaru,pines}. Detailed balance is a consequence of LTE, which, in this case, is governed by Fermi statistics where scattering events for which the electron final state falls into an energy level lower than the ground state (and consequently an energy gain for the photon) are not allowed. In an inhomogeneous system, however, currents induced by the gradients provide an additional sink or source in the energy exchange between electrons and photons. This effect is also well-known from scattering measurements of ion acoustic waves in classical plasmas in the presence of a heat flow \cite{evans}.

The change in the detailed balance relation bears important consequences in the analysis of Thomson scattering data from solid density plasmas. As discussed in Refs.~\cite{glenzer1, doppner, faustlin}, the intensity ratio of the two plasmon peaks is often assumed to provide a direct measurement of the temperature, without relying on any additional approximations. On the other hand, as discussed by Chapman and Gericke \cite{gericke}, once the homogeneous plasma or equilibrium assumption is relaxed, the intensity ratio of the plasmon peaks is no longer uniquely defined by the electron temperature. In such cases, nonequilibrium distributions and spatial gradients determine the shape and intensity of the plasmon peaks in the scattering signal. 

\begin{figure}
\centering
\includegraphics[width=0.9\linewidth]{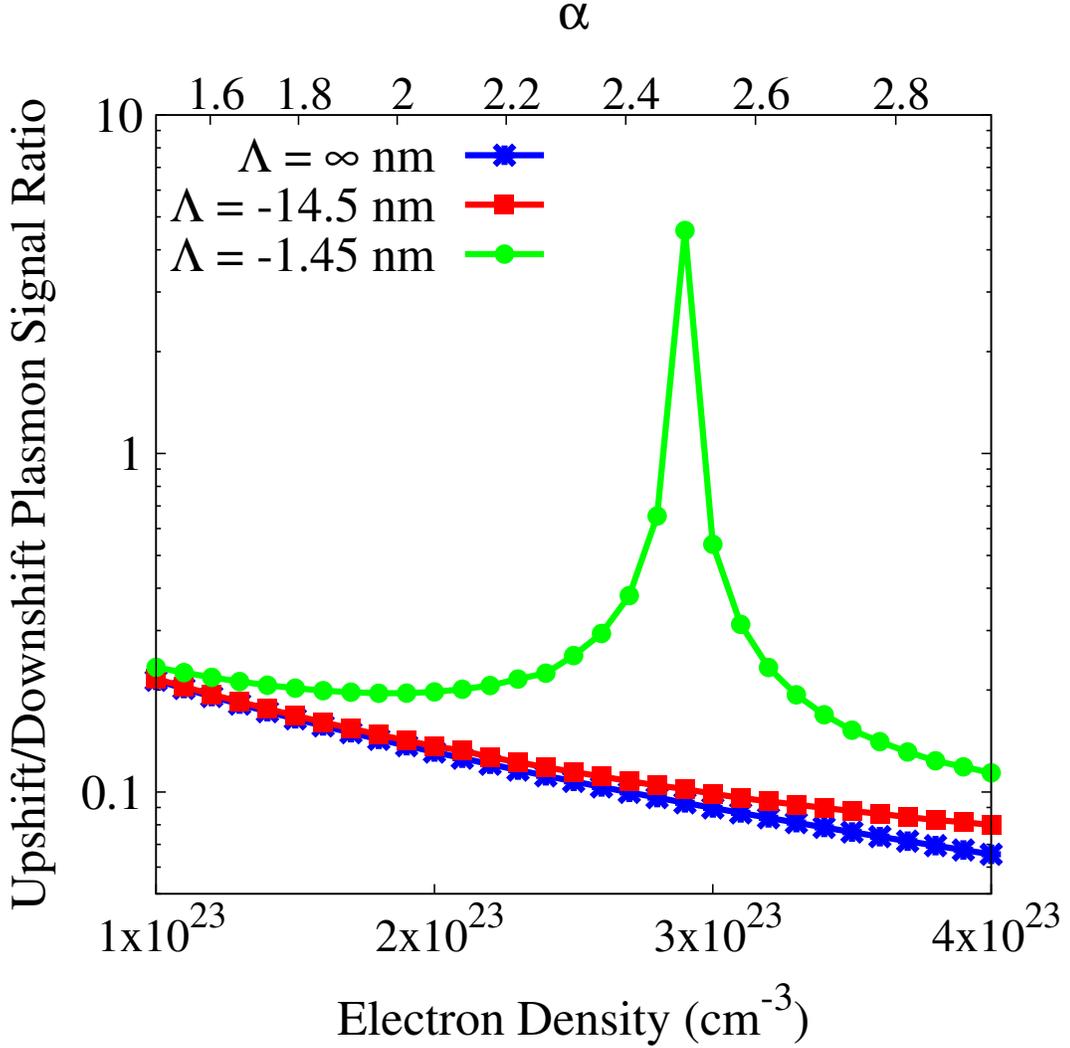}
\caption{{\bf Ratio of upshifted over downshifted plasmon signals}
The ratio of the integrated plasmon signals is given for different gradient conditions as well as the equilibrium case. The intermediate gradient (red) steadily diverges from the equilibrium case (blue) with increasing density. We see two effects for the larger gradient (green). First, the green curve begins to diverge from the equilibrium case at a lower density, as expected. Second, as density increases, the green curve peaks and subsequently decreases rapidly; this is due to the coincidental minimization of $\operatorname{Re}[{\chi^{\rm eq}\left({\bf k},\omega\right)}]$ which emphasizes the effect of the gradient expansion terms on the DSF.}
\label{fig:Ratioplasmons}
\end{figure}

Figure~\ref{fig:Ratioplasmons} demonstrates how the ratio of the upshifted and downshifted plasmon signals dramatically diverges from the equilibrium case as density varies for the same conditions given in the example above. The ratio of plasmon signals is maximized when
 $n_{e} \approx 2.8 \times 10^{23} \: \rm cm^{-3} $ which occurs when $\operatorname{Re}[{\chi^{\rm eq}\left({\bf k},\omega\right)}]$ reaches a minimum for both resonances, $\omega = \pm \omega_{pe}$, while $\dfrac{\partial}{\partial k} \operatorname{Im}[\chi^{\rm eq}\left({\bf k},\omega\right)]$ reaches a maximum for the upshifted plasmon and a minimum for the downshifted plasmon. It is due to this latter term being an odd function with respect to $\omega$ that the detailed balance condition is violated. We may conclude that given a sufficiently nonequilibrium plasma probed in the collective scattering regime, assuming the ratio of the plasmon peaks to yield a measure of temperature may lead to large errors.

\begin{figure}
\centering
\includegraphics[width=0.9\linewidth]{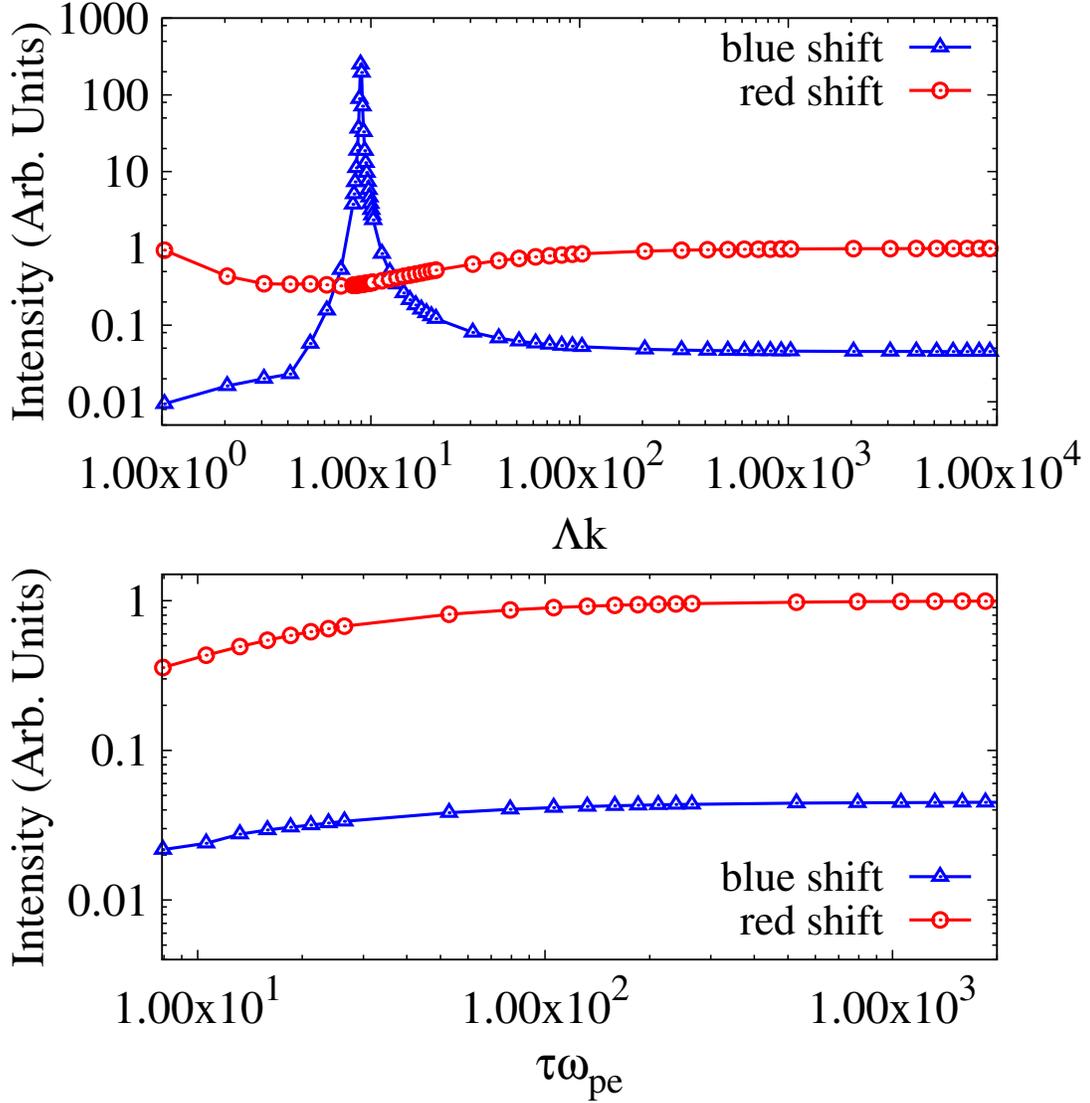}
\caption{{\bf Effects of gradients on plasmon peak intensities.} We show the variation in the peak intensities of the plasmons with respect to the gradient parameter. For spatial gradients this is $\Lambda k$ and for temporal gradients this is $\tau \omega_{pe}$ . For spatial gradients we see a sharp rise in the blue shifted plasmon and a dip in the red shifted plasmon as we approach steeper gradients ({\it i.e.}, lower gradient parameter). For temporal gradients both plasmons are reduced with higher gradients.}
\label{plasmonVary}
\end{figure}

Similarly, in Figure~\ref{plasmonVary}  we show how significantly the red and blue shifted plasmon peak intensities change as a function of the inhomogeneity scale lengths. The simulations were conducted for the same plasma parameters as discussed above, while varying either $\Lambda$ or $\tau$. Here we see that the results of the new approach deviate from equilibrium starting at around $\Lambda k \sim 100$. It is important to note, that although the figure gives a sense of how gradients may alter the scatter signal, the change is not the same for all plasma parameters. This is due to the inhomogeneous dynamic structure factor depending on differences between the real and imaginary components of the dielectric susceptibility as described above.

\subsection*{Comparison with previous approaches}

An approximate, but often applied \cite{glenzer4, rozmus, falk, chapman, sperling}, method to account for the effects of spatial gradients on the Thomson scattering spectrum is to simply take the average over a given region. Thus, individual scattering spectra are generated for different plasma regions using the FDT, weighted by the number of electrons present in the subvolume selected and finally added to obtain the full spectrum:
\begin{equation}
S_{\rm sum}({\bf k}, \omega) = \dfrac{\Sigma_{j} V_j n_{e,j} S_{j}({\bf k}, \omega)}{\Sigma_{j} V_j n_{e,j}},
\end{equation}
where $V_j$, $n_{e,j}$ and $S_{j}({\bf k}, \omega)$ are the volume, electron density and structure factor of cell $j$, respectively. 
The simplest approximation consists in taking $S_{j}({\bf k}, \omega) \equiv S^{eq}_{j}({\bf k}, \omega)$, where $S^{eq}_{j}({\bf k}, \omega)$ is the equilibrium structure factor inside the $j$th cell. This is the method discussed in Ref.~\cite{glenzer4, rozmus, falk}, for example. As already mentioned, this incoherent addition of scattering spectra is strictly valid only if $\alpha \ll 1$ or both $\Lambda/\lambda_p \gg 1$ and $\tau/\tau_p \gg 1$. One might assume that dividing the volume into infinitesimally small cells would make this method applicable even for steep gradients, however, one neglects the explicit influence of the gradients in this way. Technically, the problem with the LTE treatment is that the local dynamic structure factors which are being used assume that only the imaginary, dissipative component of the dielectric susceptibility is important, as in Eq.~\eqref{eq:DSFee}, but don't consider the current flows. This assumption fails for large gradients where the real component of the dielectric susceptibility becomes critical. Such an effect is also common in Onsager-violating media where cross-terms between the energy storing and dissipative components of the dielectric susceptibility emerge \cite{Buhmann}. Here, instead, we propose to replace the calculation of the equilibrium structure factor inside each subvolume element with its non-equilibrium version.
 
\begin{figure}[t]
\centering
\includegraphics[width=0.9\linewidth]{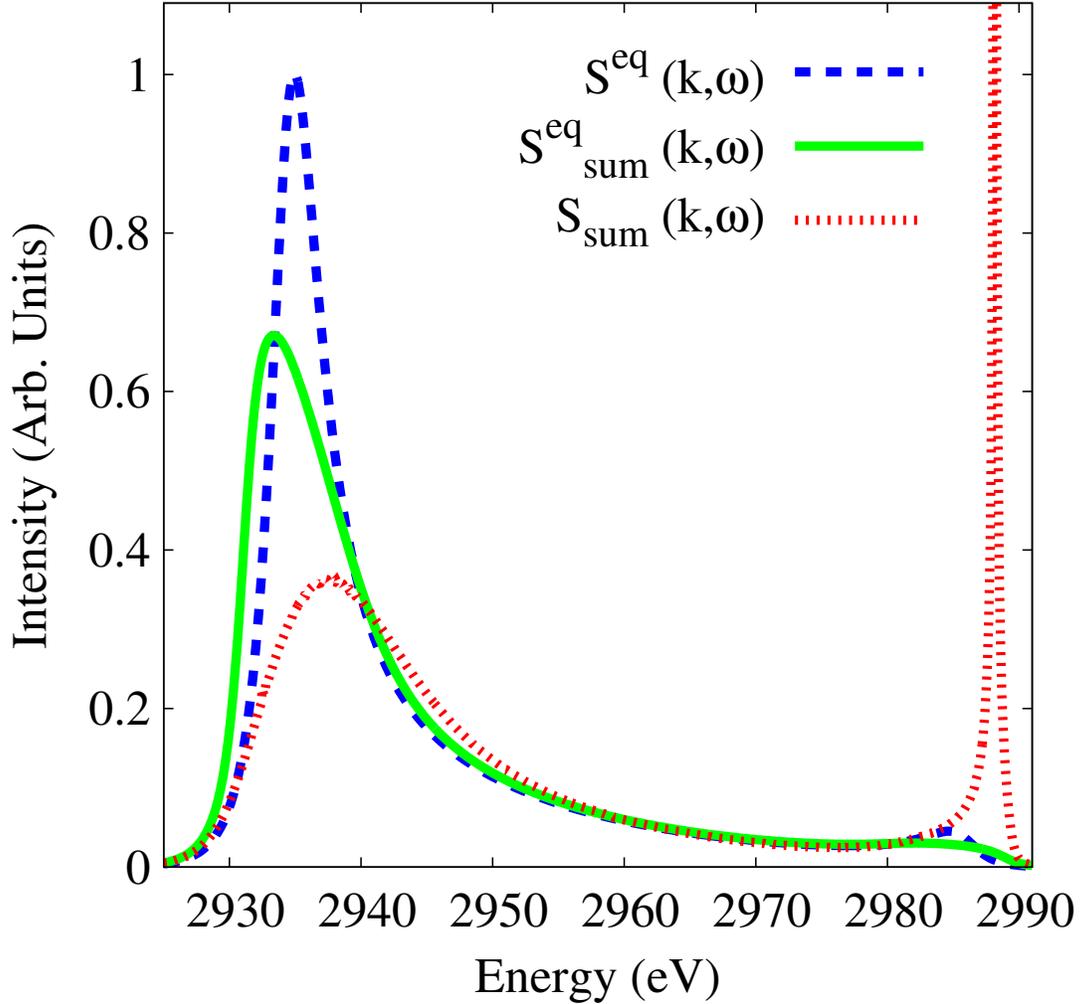}
\caption{{\bf Predicted spectra.} 
{  We have compared the homogeneous case (green line) against our full inhomogeneous scattering model (red line) using the weighted sum method in both cases and for the plasma conditions given in the text with  $\Lambda =1.45$ nm and where $k = 1.03 \times 10 \: \rm m^{-1}$. The weighted sum method was implemented by assuming a linear density profile, with $n_e$ varying from $1.1\times 10^{23}$ cm$^{-3}$ to $3.3\times 10^{23}$ cm$^{-3}$ over a distance equal to $\Lambda$. This profile is then divided into 220 equal cells and the scattering from each cell summed together. The equilibrium structure factor (assuming an homogeneous plasma) is also shown in the figure (blue line). It is clear that the upshifted plasmon peak intensity for the inhomogeneous weighted sum case is much higher than even the downshifted plasmon intensities. This plasmon peaks at an intensity of ~3 relative to the rest of the curves (not shown).}}
\label{fig:Comparisonofgradientmethods4}
\end{figure}

Let us continue our practical example and test the above approach by assuming that the plasma exhibits density variations on a scale of $\Lambda = 1.45$ nm. The density gradient is then given by $d n_e / dr = n_e/\Lambda$. We notice that such gradients are, for example, quite common in shocked dense matter. For the conditions of Ref.~\cite{regan}, we expect the electron mean free path to be $\lambda_{\rm mfp} \sim 0.4$ nm, and thus the shock width to be a few mean free paths, which is indeed of the same order as assumed here.
 Figure~\ref{fig:Comparisonofgradientmethods4} shows a comparison between the overall structure factors calculated with the methods described above. Large modifications to the whole spectrum are seen when the full effect of inhomogeneities is accounted for in the weighted sum.

\section*{Concluding Remarks}

Our work indicates that collective Thomson scattering by free electrons from strongly inhomogeneous matter requires a modeling of the dielectric response including not only the effects of microscopic, but also macroscopic spatial and temporal gradients. The treatment presented here gives results that differ significantly from those assuming uniform plasma conditions for systems with gradients as often encountered in experiments, in particular the violation of detailed balance. Further work must be done to incorporate inhomogeneities for Thomson scattering by screening clouds and ion acoustic modes, which are ubiquitous in warm dense matter and tend to be strongly correlated.

\subsection*{Acknowledgments}
This article is a tribute to the late Prof Basil Crowley, who has inspired us to be rigorous in the pursuit of knowledge. We would like to thank Prof. M. Schlanges (Greifswald) for illuminating discussions.
The research leading to these results has received funding from the European Research Council under the European Community's Seventh Framework Programme (FP7/2007-2013) / ERC grant agreement no. 256973.   

\subsection*{Author Contributions}
G.G. conceived this project. The paper was written by P.M.K, B.J.B.C and D.O.G. Theoretical guidance was given by B.J.B.C and D.O.G. The relation to experimental data was provided by S.P.R.

\subsection*{Competing Interests}
The authors declare that they have no competing financial interests.


%\nocite{*}

\end{document}